\documentclass[runningheads]{llncs}
\usepackage{graphicx}
\graphicspath{{./pics/}}
\DeclareGraphicsExtensions{.pdf,.jpeg,.png}
\usepackage{amsmath}
\usepackage{amssymb}
\usepackage{booktabs}
\usepackage{multirow}
\usepackage{color}
\usepackage[ruled,vlined]{algorithm2e}

\usepackage{algorithmic}

\begin{document}
\title{Globally Guided Progressive Fusion Network for 3D Pancreas Segmentation}
\titlerunning{Globally Guided Progressive Fusion Network}
%
\author{ Chaowei Fang \inst{1,2}  \and 
Guanbin Li\inst{3} \and 
Chengwei Pan\inst{1}  \and 
Yiming Li\inst{1} \and
Yizhou Yu\inst{1,2}
}

\authorrunning{C. Fang et al.
}
%
\institute{$^1$Deepwise AI Lab~~~~~$^2$The University of Hong Kong~~~~~$^3$Sun Yat-sen University
}
\maketitle              
\begin{abstract}
Recently 3D volumetric organ segmentation attracts much research interest in medical image analysis due to its significance in computer aided diagnosis. This paper aims to address the pancreas segmentation task in 3D computed tomography volumes. We propose a novel end-to-end network, \emph{Globally Guided Progressive Fusion Network}, as an effective and efficient solution to volumetric segmentation, which involves both global features and complicated 3D geometric information. A progressive fusion network is devised to extract 3D information from a moderate number of neighboring slices and predict a probability map for the segmentation of each slice. An independent branch for excavating global features from downsampled slices is further integrated into the network. Extensive experimental results demonstrate that our method achieves state-of-the-art performance on two pancreas datasets.

\end{abstract}
\begin{figure}[t]
\centering
\includegraphics[width=0.9\columnwidth]{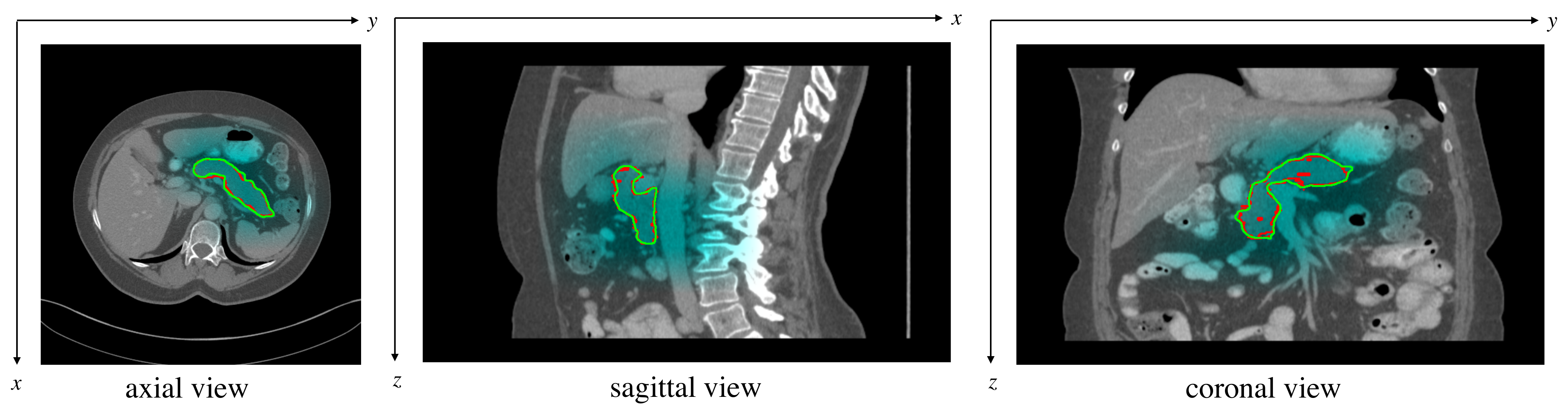}\vspace{-3mm}
\caption{An example of pancreas segmentation in the axial, sagittal and coronal views. The contour of the ground truth and our result is shown in red and green respectively. Blended regions indicate the probability map inferred from the global feature map. }
\label{fig:example}
\end{figure}

\section{Introduction}
Automatic organ segmentation, which is critical to computer aided diagnosis, is a fundamental topic in medical image analysis. 
This paper focuses on pancreas segmentation in 3D computed tomography (CT) volumes which is more difficult than segmentations of other organs such as liver, heart and kidneys~\cite{roth2015deeporgan}.

Driven by the rapid development of deep learning techniques, significant progress has been achieved on 3D volumetric segmentation~\cite{roth2018an,xia2018bridging}. State-of-the-art methods primarily fall into two categories. The first category~\cite{zhou2017a} is based on segmentation networks originally designed for 2D images, e.g. FCN~\cite{long2015fully}. However, only a small number of adjacent slices (usually 3) are stacked together as the input to take advantage of network weights pretrained on natural image datasets such as Pascal VOC~\cite{everingham2010the}. Although majority voting~\cite{zhou2016three} can be used to incorporate pseudo 3D contextual information through 2D segmentation in slices along different views, powerful 3D features are still not exploited. Methods in the other category are based on 3D convolution layers, such as V-Net~\cite{milletari2016v} and 3D U-Net~\cite{cciccek20163d,roth2018deep}. Due to the huge memory overhead of 3D convolutions, the input is either decomposed into overlapping 3D patches~\cite{cciccek20163d}, which ignores the global knowledge, or resized to a volume with a poor resolution~\cite{roth2018deep}, which likely gives rise to missed detections. Coarse-to-fine segmentation is a popular and effective choice for improving the accuracy~\cite{roth2018an,yu2018recurrent,xia2018bridging}. However, it is severely dependent on the performance of its coarse segmentation model. Omission of regions of interest (ROIs) or inaccurate size of ROIs in the coarse segmentation often lead to irreparable loss. Most of these volumetric segmentation methods have been applied in pancreas segmentation such as \cite{zhou2017a,yu2018recurrent,xia2018bridging}.

In this paper, we focus on one fixed type of organs (pancreas) and the overall spatial arrangement of organs in any human body is more or less fixed as well. In such a specialized setting, both local and global contextual information is critical for achieving highly accurate segmentation results.
To tackle the aforementioned challenges, we propose a novel end-to-end network, called \emph{Globally Guided Progressive Fusion Network}. The backbone in our method is a progressive fusion network devised to extract 3D local contextual information from a moderate number of neighboring slices and predict a 2D probability map for the segmentation of each slice. However our progressive fusion network has limited complexity and receptive fields, which are inadequate for acquiring global contextual information. Thus a global guidance branch consisting of convolution layers is employed to excavate global features from a complete downsampled slice. We elegantly integrate this branch into the progressive fusion network through sub-pixel sampling. An example of the segmentation result of our method is presented in Fig.~\ref{fig:example}. In summary, the main contributions of our paper are as follows.
{\flushleft (1)} A progressive fusion network is devised to extract 3D local contextual information from a 3D neighborhood. A unique aspect of this network is that the encoding part performs 3D convolutions while the decoding part performs 2D convolution and deconvolution operations.
{\flushleft	(2)} A global guidance branch is devised to replenish global contextual information to the progressive fusion network. The entire network, including the global branch, is trained in an end-to-end manner.
{\flushleft	(3)} Our method has been successfully validated on two pancreas segmentation datasets, achieving state-of-the-art performance.


\section{Method}
\subsection{Overview}
\begin{figure}[t]
\centering
\includegraphics[width=0.9\columnwidth]{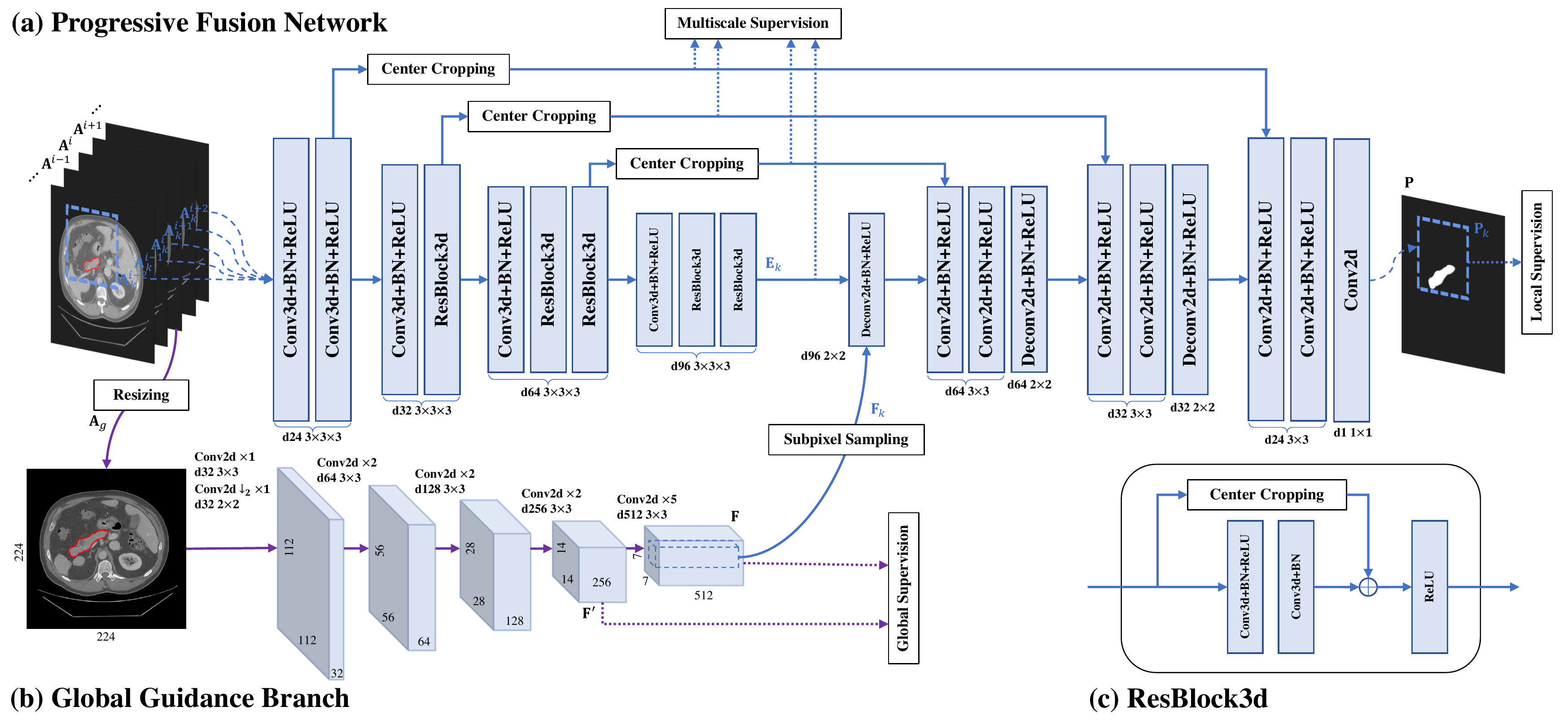}\vspace{-3mm}
\caption{The main pipeline of our method. More details are illustrated in supplemental material. (Best viewed in color)}
\label{fig:pipeline}
\end{figure}

As discussed earlier, both local and global contextual information is critical for achieving highly accurate segmentation results. On the other hand, segmentation precision, especially around boundaries, is closely related to the spatial resolution of the input volume. However the huge memory consumption of 3D volumes prevents us from loading an entire high-resolution volume at once. Considering the above factors, we devise a novel end-to-end network, which segments every slice in a patchwise manner by predicting a probability map for each 2D image patch.
This network consists of two modules: a progressive fusion network is devised to mine 3D local contextual features for a 2D image patch from its high-resolution 3D neighborhood; a global guidance branch is devised to replenish a complementary 2D global feature representation extracted from an entire downsampled slice. The overall architecture is presented in Fig.~\ref{fig:pipeline}.

Given an $l \times h\times w $ input volume, where $h$ and $w$ represent the height and width of axial slices respectively and $l$ is the number of axial slices, we define $\mathbf A^i$ ($h\times w$), $\mathbf S^i$ ($l\times h$) and $\mathbf C^i$ ($l\times w$) as the $i$-th slice in the axial, sagittal and coronal view, respectively. In the remainder of this section, we will use slices in the axial view to elaborate the aforementioned two modules. Suppose $\mathbf A^i$ is decomposed into $N$ overlapping 2D patches $\{\mathbf A^i_k| k=1,\cdots,N\}$.

\subsection{Progressive Fusion Network}\label{sec:pfn}
Local texture and shape features are valuable for organ segmentation, especially for accurate boundary localization. Hence we devise a progressive fusion network (Fig.~\ref{fig:pipeline}(a)) based on the encoder-decoder architecture to extract 3D local contextual features for each 2D image patch $\mathbf A^i_k$ from its 3D neighborhood, which includes corresponding 2D patches from a moderate number (31) of adjacent slices, $\{\mathbf A^{i+t}_k|t=-T,\cdots,T\}$. The superscript $i$ will be neglected by default for conciseness below.

The encoder, taking a 3D patch as the input, consists of 3D convolution layers and residual blocks~\cite{he2016deep}, which are organized into 4 groups. Between every two consecutive groups, max pooling is used to reduce the spatial resolution of the feature map by half, giving rise to feature maps with 4 different scales. Inspired from \cite{caballero2017real}, our network progressively fuses the slices in the input 3D patch by not performing the convolution operation in the 2 outmost slices in every 3D convolution layer because these two slices are of least relevance to the central slice. We choose $T$ to be the number of 3D convolution layers so that there exists only one slice (the central slice) in the final group of feature maps, $\mathbf E_k$. The kernel size of each convolutional layer is set to $3\times3\times3$ and the overall receptive field of the encoder is $144\times144$, only covering part of the input patch. The decoder is set up with 2D convolution and deconvolution layers, producing the final segmentation result for the central slice. As in U-Net~\cite{cciccek20163d,roth2018deep}, there exist skip connections between corresponding encoder and decoder layers. Since our encoder and decoder as well as residual blocks deal with feature maps with different dimensionality, central cropping is performed to discard surplus features in skip connections.

\subsection{Global Guidance Branch} \label{sec:global}
 Global contextual information is vital for providing absolute and relative positions with respect to distant objects. For example, the pancreas always lies in the upper center of the abdomen behind the stomach. To exploit global information, we devise a global guidance branch (Fig.~\ref{fig:pipeline}(b)) to extract a global feature map from $\mathbf A_g$ with resolution $h_g\times w_g$, which is downsampled from the original slice $\mathbf A$. This branch consists of 13 convolution layers interleaved with 4 max pooling layers. The height and width of the global feature map $\mathbf F$ is $h_g/32$ and $w_g/32$ respectively. For every pixel in the local feature map $\mathbf E_k$, sub-pixel sampling is utilized to calculate a corresponding feature vector from $\mathbf F$, resulting a global feature map $\mathbf F_k$ for $\mathbf A_k$. $\mathbf E_k$ and $\mathbf F_k$ are concatenated and fed into the decoder in the progressive fusion network.

\begin{algorithm}[t]
\algsetup{linenosize=\tiny}
\scriptsize
\caption{Inference procedure of our network.}
\KwIn{ Slices: $\mathbf A^i$, $i=1,\cdots,l$. } 
\KwOut{Probability map: $\mathbf P^i$, $i=1,\cdots,l$.}
\begin{algorithmic}[1]
\FOR{ each slice $\mathbf A$ in $\{\mathbf A^i\}$}
	\STATE Downsample $\mathbf A$ to obtain $\mathbf A_g$;
	\STATE Compute $\mathbf F$ from $\mathbf A_g$ using the global guidance branch (Section~\ref{sec:global});
	\STATE Decompose $\mathbf A$ into $N$ overlapping patches $\{\mathbf A_k|k=1,\cdots,N\}$;
	\FOR{$k=1$ to $N$}
		\STATE Sample $\mathbf F$ to obtain a global feature map $\mathbf{F}_k$ for $\mathbf A_k$;
		\STATE Extract a local feature map $\mathbf E_k$ from a 3D neighborhood of $\mathbf A_k$ using the 3D encoder of our progressive fusion network (Section \ref{sec:pfn});
		\STATE Compute the probability map $\mathbf P_k$ for $\mathbf A_k$ by feeding concatenated $\mathbf E_k$ and $\mathbf{F}_k$ through the 2D decoder in our network; 
	\ENDFOR
	\STATE Merge $\{\mathbf P_k\}$ into $\mathbf P$ after disregarding peripheral overlapped pixels;
\ENDFOR
\end{algorithmic}
\label{alg:infer}
\end{algorithm}

\subsection{Training Loss}
Let $\mathbf P$ and $\mathbf G$ be the predicted and groundtruth segmentation of the slice $\mathbf A$ respectively. $p(x,y), g(x,y)\in\{0,1\}$ indicates whether pixel ($x$, $y$) belongs to the predicted and groundtruth target region respectively. Binary cross entropy is used to measure the dissimilarity between $\mathbf P$ and $\mathbf G$,
\begin{equation}
\mathrm{C}(\mathbf P, \mathbf G)=-\frac{1}{wh}\sum_{x=0}^{w-1}\sum_{y=0}^{h-1} g(x,y)\log p(x,y)+(1- g(x,y))\log (1-p(x,y)).
\end{equation}

We also use a fully connected layer to predict a probability map for each scale of the feature maps in the encoder. Let $\mathbf P^{(j)}_k$ be the probability map computed from the last feature map in the $j$-th scale. Multiscale supervision is imposed on these probability maps to enhance the training of the encoder. Likewise we also use $\mathbf F$ and the second last scale of feature $\mathbf F'$ to infer probability maps $\mathbf P^f$ and $\mathbf P^{f'}$ respectively, then impose additional supervision on the global guidance branch. The overall loss function can be summarized as follows,
\begin{equation}
\mathrm L= \frac{1}{N}\sum_{k=1}^N[\mathrm{C}(\mathbf P_k,\mathbf G_k)+\frac{1}{4}\sum_{j=1}^4\mathrm{C}(\mathbf P^{(j)}_k,\mathbf G^{(j)}_k)] +\alpha \mathrm{C}(\mathbf P^f,\mathbf G^f)+\beta \mathrm{C}(\mathbf P^{f'},\mathbf G^{f'}),
\end{equation}
where $\alpha$ and $\beta$ are constants; $\mathbf G_k$, $\mathbf G^{(j)}_k$, $\mathbf G^f$ and $\mathbf G^{f'}$ are ground truths; $\mathbf G^{(j)}_k$ is downsampled from $\mathbf G_k$; $\mathbf G^f$ and $\mathbf G^{f'}$ are downsampled from the full resolution ground truth of $\mathbf A_g$.

\paragraph{\bf The inference procedure} is summarized in Algorithm~\ref{alg:infer}. The same algorithm is applied to the segmentation of the slices from the sagittal and coronal views. The results for all three views are fused through weighted averaging~\cite{zhou2016three} to produce the pseudo-3D segmentation result. Let the predictions for the axial, sagittal and coronal views are $\mathbf V_a$, $\mathbf V_s$ and $\mathbf V_c$ respectively. The final result is $\mathbf V=w_a\mathbf V_a+w_s\mathbf V_s+w_c\mathbf V_c$, where $w_a$, $w_s$ and $w_c$ are constants.

\section{Experiments}
\subsection{Datasets}
Two pancreas datasets are used to validate the performance of the proposed 3D volumetric segmentation algorithm in this paper.
\begin{itemize}
	\item[(1)] \textbf{MSD} (short for Medical Segmentation Decathlon challenge) provides 281 volumes of CT with labelled pancreas mask. The spatial resolution is $512\times 512$ and the number of slices varies from 37 to 751. We randomly split them into 236 volumes for training, 5 for validation and 40 for testing.
	\item[(2)] \textbf{NIHC~\cite{roth2015deeporgan}} contains 82 abdominal contrast enhanced 3D CT scans with the spatial resolution equal to $512\times512$ pixels and the number of slices falling between 181 and 466. 
	We randomly split them into 48 volumes for training, 5 for validation and 29 for testing.
\end{itemize}
To measure the performance of segmentation algorithms, we first threshold the segmentation probability map by 0.5. Then Dice similarity coefficient (DSC) is used to calculate the similarity between the predicted segmentation mask and the ground truth.
\subsection{Implementation}
Because a patient's pancreas only occupies a small percentage of voxels in a CT volume, we use the following strategy to balance positive and negative training samples: two patches are cropped out from all slices of each volume; the central point of the first patch is randomly chosen from the whole volume while that of the second patch is randomly chosen from the box encompassing the pancreas. Random rotation and elastic deformation are applied to augment the training samples. The patch size is set to  $256\times 256$ for all views of NIHC and axial view of MSD. For the sagittal and coronal views of MSD, $128\times 256$ patch size is utilized. The same patch size is used in validation and the number of overlapping pixels is set to 64. The global guidance branch is trained alone for 1000 epochs using a batch size of 32 and $\alpha=\beta=0.5$. The progressive fusion network is also trained alone for 1000 epochs. Then the whole network is fine-tuned for another 800 epochs with $\alpha=0.01$ and $\beta=0$. We adopt a batch size of 4 in the latter two stages. The training process takes around 60 hours. Adam is adopted to optimize network parameters with learning rate of $10^{-4}$. The model achieving the best performance on the validation set is chosen as the final version.

\paragraph{\bf Parameters} In MSD, the difficulty of segmenting the sagittal and coronal slices is higher than segmenting axial slices as the resolution along the $z$ axis varies much. We empirically set $w_a=0.8$, $w_s=0.1$ and $w_c=0.1$ for MSD. $w_a$, $w_s$ and $w_c$ are set as $1/3$ for NIHC. $h_g$ and $w_g$ are set to 224 except for the sagittal and coronal views in MSD where 128 is used for $h_g$. $N$ is set to 1 during testing.


\subsection{Experimental Results}
\begin{table*}[t]
\caption{Comparisons with state-of-the-art segmentation algorithms.}\label{tab:comp}
\centering
\fontsize{8pt}{10pt}
\selectfont
\setlength\tabcolsep{3pt}
\begin{tabular}{ l|l|c|c|l|c|c|c }\specialrule{.1em}{0em}{0em}
  \multicolumn{1}{c|}{\multirow{2}{*}{method}} &  \multicolumn{3}{c|}{MSD} & \multicolumn{3}{c|}{NIHC}
 &\multirow{2}{*}{\#Params} \\ \cline{2-7}
& mean$\pm$std & min & max & mean$\pm$std & min & max &  \\ \hline
3D Unet-Patch\cite{roth2018an}
& 79.98$\pm$7.71 & 61.14  & 93.73 & 78.36$\pm$13.04 & 23.93  & 90.25 & $1.9\times10^7$  \\
3D Unet-Full\cite{roth2018deep}
& 81.13$\pm$8.20 & 61.84  & 93.49 & 81.43$\pm$7.53   & 49.36  & 89.60 & $1.3\times10^7$\\ \hline
2D FCN8s-A\cite{long2015fully}
& 82.24$\pm$6.88 & 62.99  & 92.61 & 81.35$\pm$5.87  & 60.57  & 88.16 & $1.3\times10^8$ \\
2D RSTN-A\cite{yu2018recurrent}
& 83.29$\pm$6.58 & \textbf{66.23}  & 92.40 & 82.56$\pm$5.18  & 63.36  & 89.82 & $2.7\times10^8$ \\
2D GGPFN-A
& 84.56$\pm$7.95 & 59.41  & 95.29 & 83.71$\pm$5.83  & 66.33  & 90.13 & $1.4\times10^7$  \\ \hline
P3D FCN8s\cite{zhou2016three}
& 82.52$\pm$7.00 & 61.75  & 92.86 & 83.24$\pm$5.63 & 61.53  & 90.13 & $4.0\times10^8$ \\

P3D RSTN\cite{yu2018recurrent}
& 83.63$\pm$6.65 & 64.21  & 93.02 & 84.45$\pm$4.89 & 66.47  & 90.80 & $8.1\times10^8$ \\
P3D GGPFN
& \textbf{84.71$\pm$7.13} & 58.62  & \textbf{95.54} &  \textbf{85.46}$\pm$\textbf{4.80} & \textbf{67.03}  & \textbf{92.24} & $4.2\times10^7$ \\ \specialrule{.1em}{0em}{0em}
\end{tabular}
\end{table*}

\begin{table}[t]
\caption{Ablation study on MSD.}\label{tab:ablation}
\centering
\fontsize{8pt}{10pt}
\selectfont
\setlength\tabcolsep{8pt}
\begin{tabular}{ c|c|c|l|c|c }\specialrule{.1em}{0em}{0em}
global guidance & 3D fusion mode & $T$ & mean$\pm$std & min & max \\ \hline
 $\checkmark$ & one-off & 1 & 78.56$\pm$8.63 & 58.76  & 93.62 \\
 $\checkmark$ & one-off & 5 & 79.62$\pm$7.65 & \textbf{60.01}  & 93.63 \\
 $\checkmark$ & one-off & 10 & 77.30$\pm$8.38 & 59.21  & 92.69 \\
 $\checkmark$ & one-off & 15 & 76.96$\pm$9.38 & 57.67  & 94.26 \\ \hline
 $\checkmark$ & progressive & 5 & 80.30$\pm$8.41 & 49.30  & 93.48 \\
 $\checkmark$ & progressive & 10 & 83.34$\pm$7.90 & 54.38  & 94.70 \\
 $\times$     & progressive & 15 & 83.46$\pm$8.15 & 56.94  & 94.28 \\
 $\checkmark$ & progressive & 15 & \textbf{84.56$\pm$7.95} & 59.41 & \textbf{95.29} \\ \hline
\specialrule{.1em}{0em}{0em}
\end{tabular}
\end{table}

\subsubsection{Comparisons with State-of-the-Art Segmentation Algorithms} 
Comparisons against state-of-the-art volumetric segmentation algorithms are reported in Table~\ref{tab:comp}. According to output type, we classify them into three categories: 3D models which predict 3D probability maps directly (such as UNet-Patch~\cite{roth2018an} and UNet-Full~\cite{roth2018deep}), 2D models which produce 2D segmentation results over slices in the axial view (such as FCN8s~\cite{long2015fully}), Pseudo-3D (P3D) models which fuse 2D segmentation results for axial, sagittal and coronal views (such as RSTN~\cite{yu2018recurrent}). Our globally guided progressive fusion network (GGPFN) can be easily integrated into the 2D and P3D segmentation frameworks.  All models used for comparison here are retrained with the datasets adopted in this paper.
Our method consistently performs better than FCN8s and RSTN in both 2D and P3D segmentation frameworks. For example, in the 2D framework, the mean DSC of our model is clearly higher than that of RSTN. With the help of the P3D segmentation framework, our algorithm achieves the best performance among all considered algorithms. Comparisons of precision-recall curves are presented in supplemental material.

\subsubsection{Ablation Study}
To demonstrate the efficacy of our globally guided progressive fusion network, we conduct an ablation study (Table~\ref{tab:ablation}) on the testing set of the MSD dataset using slices along the axial view. We implement an one-off fusion mode, which directly fuses multiple adjacent slices into a single slice by using a single convolution layer and treating the multiple slices as channels of a single slice fed into this convolution layer. Our progressive fusion mode is able to make use of 3D information more effectively. As more slices are used, the advantages of our progressive fusion network become more prominent while the one-off mode fails to discover additional useful information when the number of slices exceeds 21. The feature map produced by the global guidance branch is also able to improve segmentation performance. The mean DSC is decreased by 0.011 when the global guidance branch is disabled.
\begin{figure}[t]
\centering
\includegraphics[width=0.9\columnwidth]{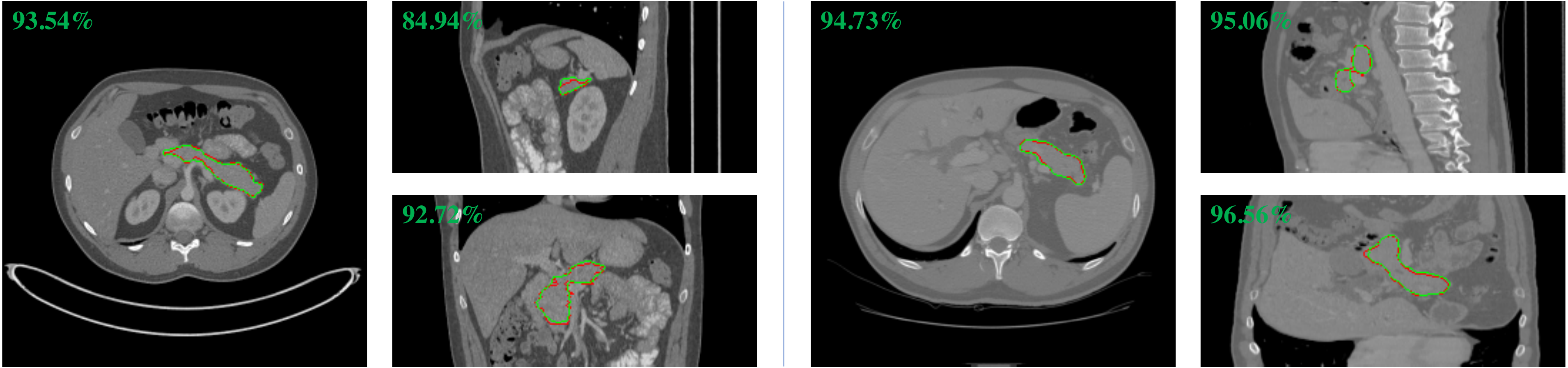}\vspace{-3mm}
\caption{Visualizations of segmentation results (green contours) produced by our method. The number on the top-left corner of each image indicate DSC metric. }
\label{fig:exper-vis}
\end{figure}

Two examples of segmented pancreas organs using our method are visualized in Fig.~\ref{fig:exper-vis}. More results are shown in supplemental material.
\section{Conclusions}
In this paper, we have presented a novel end-to-end network for 3D pancreas segmentation. The proposed network consists of a progressive fusion network and a global guidance branch. Our new algorithm achieves state-of-the-art performance on two benchmark datasets. In our future work, we will extend the application of our algorithm to multi-organ segmentation scenes and improve its boundary locating capability.


\bibliographystyle{splncs04}
\bibliography{ggpfn}

\begin{thebibliography}{10}
\providecommand{\url}[1]{\texttt{#1}}
\providecommand{\urlprefix}{URL }
\providecommand{\doi}[1]{https://doi.org/#1}

\bibitem{caballero2017real}
{Caballero}, J., {Ledig}, C., {Aitken}, A.P., {Acosta}, A., {Totz}, J., {Wang},
  Z., {Shi}, W.: Real-time video super-resolution with spatio-temporal networks
  and motion compensation. In: 2017 IEEE Conference on Computer Vision and
  Pattern Recognition (CVPR). pp. 2848--2857 (2017)

\bibitem{cciccek20163d}
{\c{C}}i{\c{c}}ek, {\"O}., Abdulkadir, A., Lienkamp, S.S., Brox, T.,
  Ronneberger, O.: 3d u-net: learning dense volumetric segmentation from sparse
  annotation. In: International conference on medical image computing and
  computer-assisted intervention. pp. 424--432. Springer (2016)

\bibitem{everingham2010the}
{Everingham}, M., {Gool}, L.J.V., {Williams}, C.K.I., {Winn}, J.M.,
  {Zisserman}, A.: The pascal visual object classes (voc) challenge.
  International Journal of Computer Vision  \textbf{88}(2),  303--338 (2010)

\bibitem{he2016deep}
{He}, K., {Zhang}, X., {Ren}, S., {Sun}, J.: Deep residual learning for image
  recognition. In: 2016 IEEE Conference on Computer Vision and Pattern
  Recognition (CVPR). pp. 770--778 (2016)

\bibitem{long2015fully}
{Long}, J., {Shelhamer}, E., {Darrell}, T.: Fully convolutional networks for
  semantic segmentation. In: 2015 IEEE Conference on Computer Vision and
  Pattern Recognition (CVPR). pp. 3431--3440 (2015)

\bibitem{milletari2016v}
{Milletari}, F., {Navab}, N., {Ahmadi}, S.A.: V-net: Fully convolutional neural
  networks for volumetric medical image segmentation. In: 2016 Fourth
  International Conference on 3D Vision (3DV). pp. 565--571 (2016)

\bibitem{roth2015deeporgan}
{Roth}, H.R., {Lu}, L., {Farag}, A., {Shin}, H.C., {Liu}, J., {Turkbey}, E.B.,
  {Summers}, R.M.: Deeporgan: Multi-level deep convolutional networks for
  automated pancreas segmentation. medical image computing and computer
  assisted intervention pp. 556--564 (2015)

\bibitem{roth2018an}
{Roth}, H.R., {Oda}, H., {Zhou}, X., {Shimizu}, N., {Yang}, Y., {Hayashi}, Y.,
  {Oda}, M., {Fujiwara}, M., {Misawa}, K., {Mori}, K.: An application of
  cascaded 3d fully convolutional networks for medical image segmentation.
  Computerized Medical Imaging and Graphics  \textbf{66},  90--99 (2018)

\bibitem{roth2018deep}
{Roth}, H.R., {Shen}, C., {Oda}, H., {Oda}, M., {Hayashi}, Y., {Misawa}, K.,
  {Mori}, K.: Deep learning and its application to medical image segmentation.
  Medical imaging technology  \textbf{36}(2),  63--71 (2018)

\bibitem{xia2018bridging}
{Xia}, Y., {Xie}, L., {Liu}, F., {Zhu}, Z., {Fishman}, E.K., {Yuille}, A.L.:
  Bridging the gap between 2d and 3d organ segmentation with volumetric fusion
  net. medical image computing and computer assisted intervention pp. 445--453
  (2018)

\bibitem{yu2018recurrent}
{Yu}, Q., {Xie}, L., {Wang}, Y., {Zhou}, Y., {Fishman}, E.K., {Yuille}, A.L.:
  Recurrent saliency transformation network: Incorporating multi-stage visual
  cues for small organ segmentation. In: 2018 IEEE/CVF Conference on Computer
  Vision and Pattern Recognition. pp. 8280--8289 (2018)

\bibitem{zhou2016three}
{Zhou}, X., {Ito}, T., {Takayama}, R., {Wang}, S., {Hara}, T., {Fujita}, H.:
  Three-dimensional ct image segmentation by combining 2d fully convolutional
  network with 3d majority voting. International Workshop on Large-Scale
  Annotation of Biomedical Data and Expert Label Synthesis pp. 111--120 (2016)

\bibitem{zhou2017a}
{Zhou}, Y., {Xie}, L., {Shen}, W., {Wang}, Y., {Fishman}, E.K., {Yuille}, A.L.:
  A fixed-point model for pancreas segmentation in abdominal ct scans. medical
  image computing and computer assisted intervention pp. 693--701 (2017)

\end{thebibliography}

\newpage
\section{Supplementary Material}
\subsection{Precision-recall Curves}
\begin{figure}[h]
\centering
\begin{minipage}{\textwidth}
\centering
\includegraphics[width=0.45\textwidth]{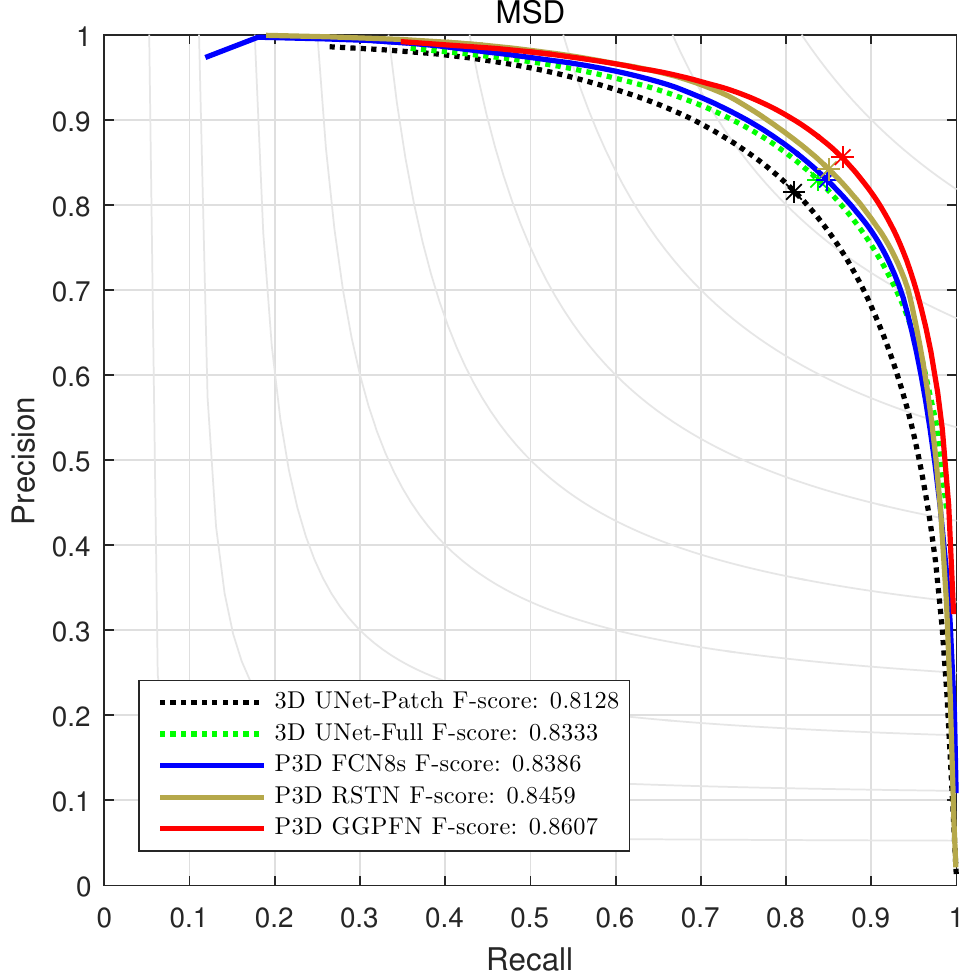} 
\includegraphics[width=0.45\textwidth]{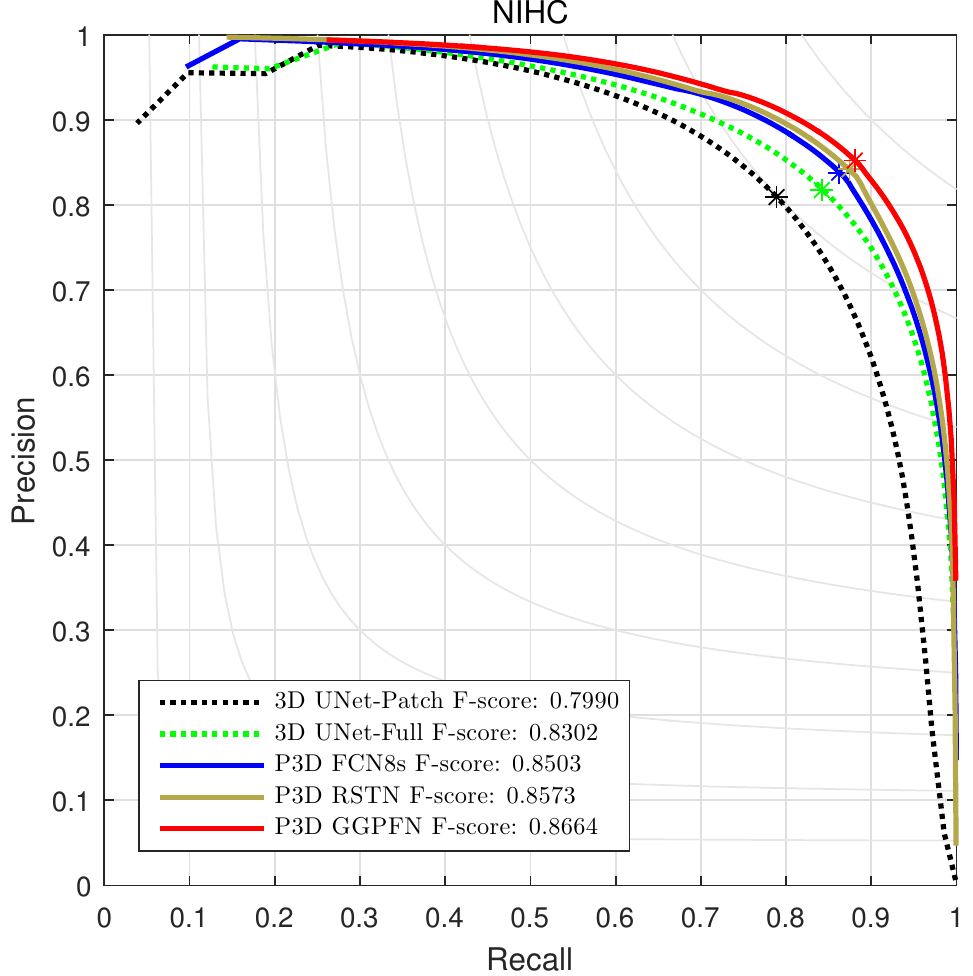} 
\end{minipage}
\caption{Precision-recall curves and F-scores in two datasets. }
\label{fig:prcurve}
\end{figure}

Precision-recall curves and F-scores of 3D UNet-Patch \cite{roth2018an}, 3D UNet-Full \cite{roth2018deep}, P3D FCN8s \cite{zhou2016three}, P3D RSTN \cite{yu2018recurrent} and our final model P3D GGPFN are presented in Fig.~\ref{fig:prcurve}. Our method shows superiority to other methods.
\begin{figure}
\centering
\includegraphics[width=0.9\textwidth]{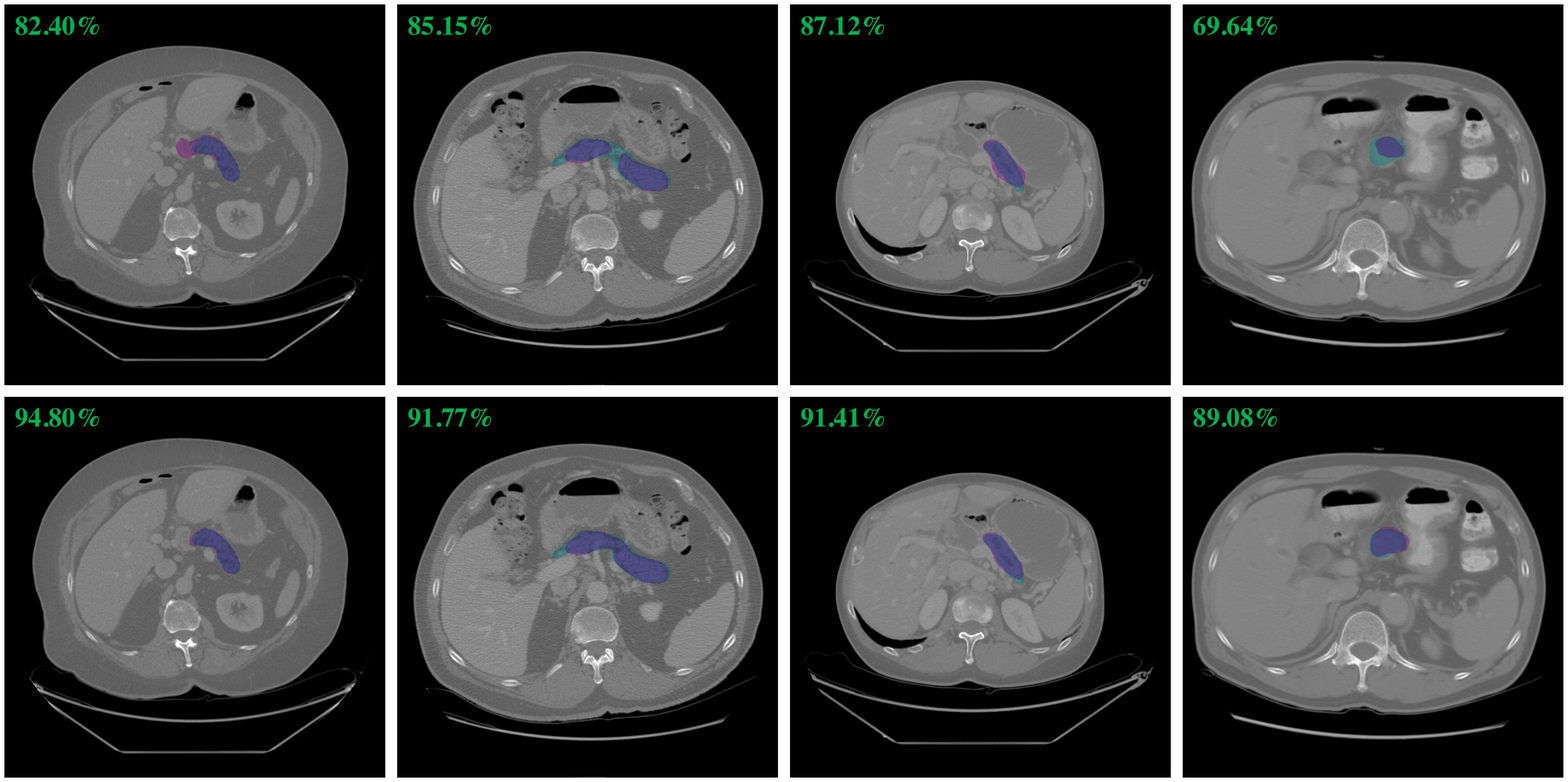} \vspace{-15mm}
\caption{Comparison of our method with or without global feature. The first row shows results of our method without global guidance branch. The second row presents results of our model using global guidance branch. True positive, false positive and false negative regions are shown in blue, green and red respectively.}\label{fig:cmp-global}
\end{figure}

\begin{figure}
\centering
\includegraphics[width=0.9\textwidth]{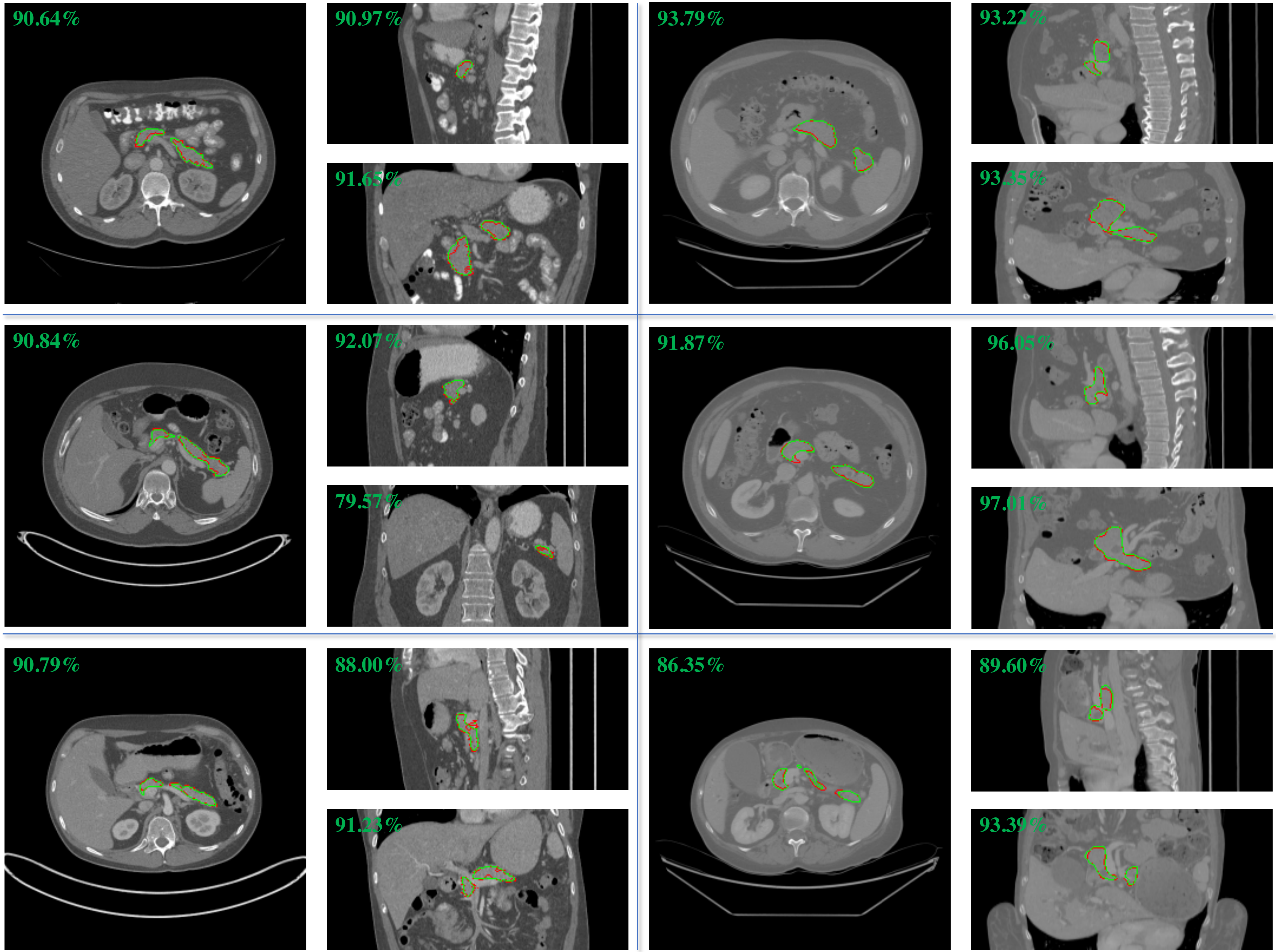}\vspace{-3mm}
\caption{More examples of segmentation results generated by our method. The contour of groundtruth is shown in red and the contour of segmentation result is shown in green.}\label{fig:exper-vis}
\end{figure}
\subsection{Qualitative Results}
Comparison of our method with or without global feature is shown in Fig. \ref{fig:cmp-global}. More visualizations of segmentation results produced by our proposed method are shown in Fig. \ref{fig:exper-vis}.

\end{document}